\documentclass{article}
\usepackage{spconf,amsmath,graphicx,color,amsfonts,bm,cite,url}
\usepackage[export]{adjustbox}

\newcommand{\setC}{\ensuremath{\mathbb{C}}}
\newcommand{\setR}{\ensuremath{\mathbb{R}}}

\title{ChannelAugment: Improving generalization of multi-channel ASR by training with input channel randomization}
\name{Marco Gaudesi$^*$\thanks{$^*$Equal contribution.}, Felix Weninger$^*$, Dushyant Sharma, Puming Zhan}
\address{Nuance Communications}
\begin{document}
\ninept
\maketitle
\begin{abstract}
End-to-end (E2E) multi-channel ASR systems show state-of-the-art performance in far-field ASR tasks by joint training of a multi-channel front-end along with the ASR model.
The main limitation of such systems is that they are usually trained with data from a fixed array geometry, which can lead to degradation in accuracy when a different array is used in testing.
This makes it challenging to deploy these systems in practice, as it is costly to retrain and deploy different models for various array configurations.
To address this, we present a simple and effective data augmentation technique, 
which is based on randomly dropping channels in the multi-channel audio input during training, in order to improve the robustness to various array configurations at test time.
We call this technique ChannelAugment, in contrast to SpecAugment (SA) which drops time and/or frequency components of a single channel input audio.
We apply ChannelAugment to the Spatial Filtering (SF) and Minimum Variance Distortionless Response (MVDR) neural beamforming approaches. 
For SF, we observe 10.6\,\% WER improvement across various array configurations employing different numbers of microphones.
For MVDR, we achieve a 74\,\% reduction in training time without causing degradation of recognition accuracy.
\end{abstract}
\section{Introduction}
\label{sec:intro}

In recent years, the end-to-end (E2E) modeling technique, which  uses a single neural network model to subsume acoustic, pronunciation, and language modeling  \cite{Graves2012-STW,Chan2016-LAA,Weninger2019-LAS,Zeyer2019-ACO,Zhang2020-TTA,Gulati2020-CCA,Tuske2020-SHA}, has led to significant accuracy gains for ASR systems. 
In E2E multi-channel ASR systems, the neural network model also includes the multi-channel signal processing and feature extraction part \cite{Sainath2017-MSP}.
The multi-channel frontend is often realized by neural beamforming techniques \cite{Heymann2016-NNB,Erdogan2016-IMB,Drude2019-UTO}.
In contrast to the classical beamforming methods such as \cite{Doclo2007-SBR,Souden2010-OOF}, the neural beamformer can be trained jointly with the E2E model to maximize ASR accuracy \cite{Li2017-AMF,Ochiai2017-UAF,Chang2019-MSE}. 
In order to reach state-of-the-art accuracy, large amounts of training data are required for training E2E ASR systems.
Thus, Data Augmentation (DA) techniques, which can generate a virtually infinite amount of training data by randomly perturbing the training examples without altering the labels, are often used in training such models \cite{Saon2019-SNI,Nguyen2020-IST}.
In particular, the SpecAugment (SA) approach proposed in \cite{Park2019-SAA} has shown impressive improvement for E2E ASR systems in both accuracy and noise robustness \cite{Park2020-SOL}.
However, to our knowledge, few studies so far have applied data augmentation to the multi-channel ASR scenario.
Examples include speed/volume perturbation \cite{Schrank2016-DBA}, MIMO source separation \cite{Fujita2016-DAU} or white noise injection \cite{Yalta2019-CBM}.

In contrast to earlier works, our paper proposes data augmentation to address the robustness of E2E multi-channel ASR to mismatched array configurations.
The main limitation of many neural beamforming algorithms (such as \cite{Li2017-AMF,Wu2019-FDM,Park2020-RMC}) is that they assume matched array configurations in training and test, 
making it hard to deploy a single model that works with different array geometries.
To address this issue, we introduce ChannelAugment, a generic DA technique which randomly varies the set of input channels during training to improve the generalization to different array configurations.
We demonstrate the effectiveness of the technique by applying the Spatial Filtering (SF) neural beamforming algorithm \cite{Wu2019-FDM} in a variety of linear arrays employing different numbers of sensors, on top of a strong multi-channel E2E ASR baseline that already employs SA and dropout.

Furthermore, the random dropping of input channels can also reduce the training time required for the E2E ASR system. 
As the computational cost and memory consumption for neural beamforming grows linearly with the number of input channels, training on the full set of channels does not necessarily yield the optimal cost/accuracy trade-off.
In our paper, we show that ChannelAugment can greatly reduce the training time required for E2E ASR using the neural Minimum Variance Distortionless Response (MVDR) frontend \cite{Ochiai2017-UAF}, while preserving the accuracy across different array configurations.

\section{Related Work}
\label{sec:relatedwork}

In \cite{Kumatani2019-MGS}, the problem of training/test mismatch in terms of array geometry was addressed by using multi-condition training to learn various array geometries in a SF layer. However, this work did not vary the number of channels in training, did not consider training efficiency, and the method was not applied to E2E ASR. 
In \cite{Stolcke2011-MTM}, the author proposed a technique called "leave-one-out beamforming", which drops one channel at a time in a round-robin manner before beamforming, to generate multiple hypotheses during recognition for doing confusion network combination. The same technique was then extended in \cite{Yoshioka2019-MTU} by using multiple asynchronous distant microphones from different devices, but was still applied only at recognition time.
In \cite{Ochiai2017-UAF}, the neural MVDR beamformer was evaluated using various numbers of channels, without modifying the training procedure as in our work.
In \cite{Kovacs2019-ETC,Kovacs2017-ITR}, a technique named channel dropout is introduced, which however drops out frequency bands in a single-channel signal similar to SpecAugment, not channels in a multi-channel frontend as in our work.
In \cite{Shimada2020-SEL}, a random channel dropping technique is used in training, however it is not applied to an ASR system nor to a beamforming technique.
In \cite{Mosner2020-UVD} audio channels are dropped with the aim of augmenting data for training a speaker verification system with beamforming as a frontend.
The novelty of our paper is data augmentation by dropping out channels from multi-channel input in E2E ASR training, which was not addressed in previous work to our knowledge. 

\section{Methodology}
\label{sec:methodology}


\subsection{End-to-End Multi-Channel ASR}

The proposed E2E ASR model architecture is based on a multi-channel frontend (cf.\ \cite{Ochiai2017-UAF,Wu2019-FDM}) and an encoder-decoder structure with attention  (cf.\ \cite{Chan2016-LAA,Weninger2020-SSL}).
The far-field ASR task is treated as a sequence-to-sequence learning problem:
The model $M$ is trained to predict a sequence  of symbols $y_j$ (here, we use sub-word units) from the multi-channel complex spectrum $\bm{X} \in \setC^{T \times F \times C}$, where $T$ is the number of frames, $F$ is the number of frequency bins, and $C$ is the number of channels in a input utterance.
During training, $\bm{X}$ is dynamically modified by the ChannelAugment method proposed in Section \ref{sec:CA}.
The multi-channel frontend yields a single-channel enhanced power spectrum $\hat{\bf X} \in \setR^{T \times F}$.
In our work, this is achieved by the SF (Section \ref{sec:sf}) or MVDR (Section \ref{sec:mvdr}) neural beamformers.
Then, ASR features $\tilde{\bf X}$ are computed from the enhanced power spectrum.
We use a log-Mel transformation with fixed weight and apply per-utterance cepstral mean normalization as well as batch normalization. 
After this, we apply SpecAugment (Section \ref{sec:SA}) as a second stage of data augmentation to the normalized features.

From the (augmented) ASR features, the encoder $e$ creates a hidden representation.
In our work, $e$ is implemented as a stack of convolutional layers followed by bidirectional Long Short-Term Memory (bLSTM) layers, which also perform frame decimation.
The decoder takes the context vector $c_j$, which is computed by using Bahdanau attention \cite{Bahdanau2015-NMT}, as input to generate the output probability distribution $p_j = p(y_j | y_1, \dots, y_{j-1}, x) = M(x,y_{<j})$ for the $j$-th symbol.

\begin{figure}
    \centering
    \includegraphics[width=.6\columnwidth]{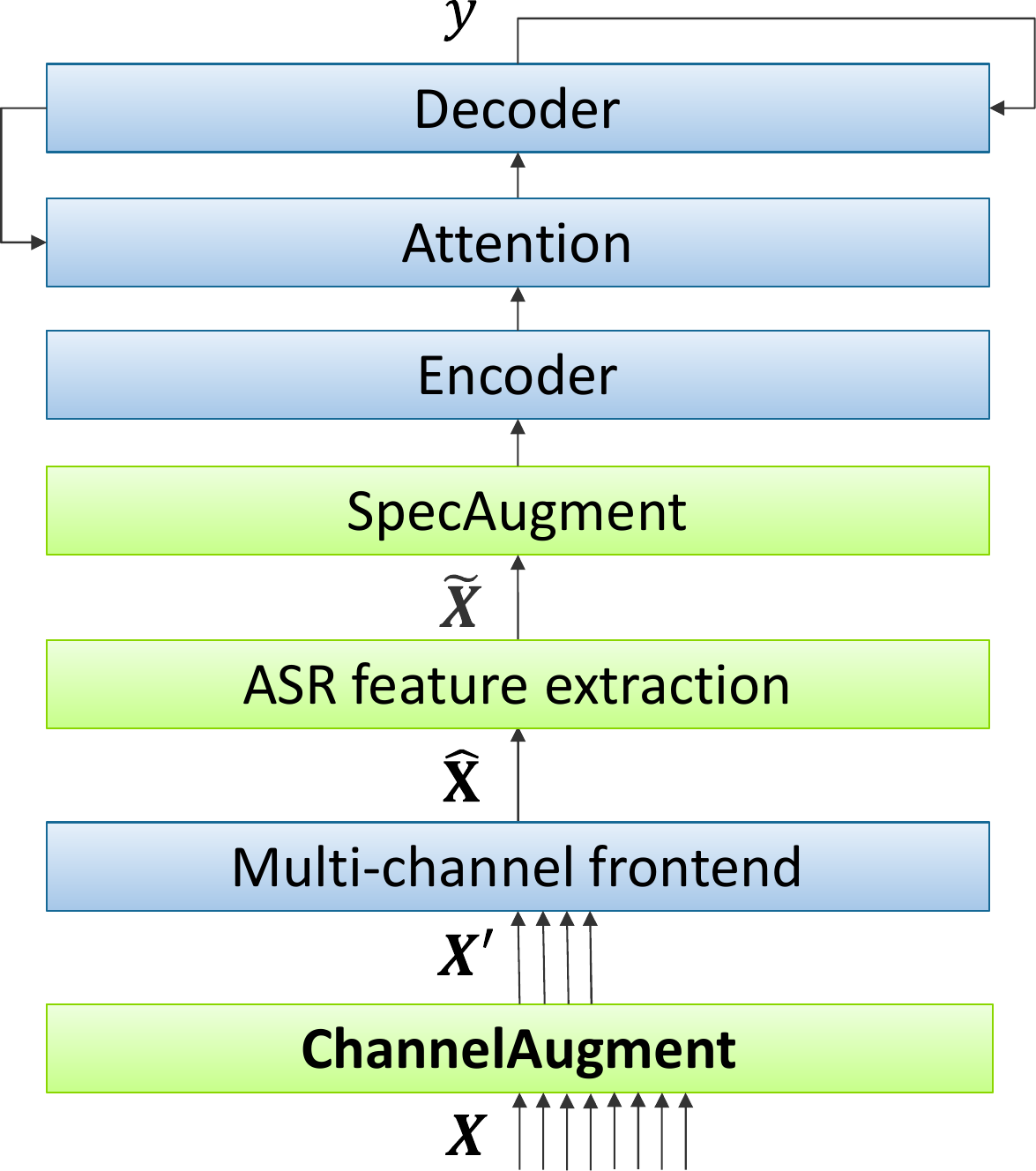}
    \caption{E2E multi-channel ASR model training with two-stage data augmentation by the proposed ChannelAugment technique as well as SpecAugment.}
    \vspace{-3mm}
    \label{fig:architecture}
\end{figure}

To perform ASR using the E2E model, hypotheses are generated by beam search.
The sequence $y_1, \dots, y_j, \dots$ of output symbols is produced one at a time, until the end-of-sentence symbol is reached. 
Figure~\ref{fig:architecture} shows the architecture of our E2E multi-channel ASR model.

\subsection{Spatial Filtering Layer}

\label{sec:sf}

The SF layer \cite{Wu2019-FDM,Kumatani2019-MGS,Park2020-RMC} is a complex-valued hidden layer designed to mimic a filter-and-sum beamforming operation with time-invariant beamforming coefficients.
The input is the multi-channel complex spectrum $\bm{X} \in \setC^{T \times F \times C}$.
The SF layer computes a tensor $\bm{Y} \in \setC^{T \times F \times D}$ as follows:
\begin{equation}
    y_{t,f,d} = \sum_c w_{f,d,c} \, x_{t,f,c} + b_{f,d},
    \label{eq:sf}
\end{equation}
where $D$ is the number of look directions of the beamformer, $t$, $f$, $d$ and $c$ are the time, frequency, look direction and input channel indices, $w_{f,d,c} \in \setC$ is a trainable weight, $x_{t,f,c}$ is an STFT coefficient of the time-frequency bin $(t,f)$ of the $c$-th channel of the input signal, and $b_{f,d} \in \setC$ is a bias term.
After the SF layer, an average pooling layer is inserted to compute the enhanced power spectrogram $\hat{\bf X} \in \setR^{T \times F}$: 
\begin{equation}
    \hat{x}_{t,f} = \frac{1}{D} \sum_d | y_{t,f,d} |^2 .
\end{equation}

\subsection{Neural MVDR}

\label{sec:mvdr}

The Neural MVDR \cite{Ochiai2017-UAF} is a frequency-domain beamformer that estimates the optimal
filter coefficients for each input utterance.
As in the SF approach, the input is the multi-channel complex-valued spectrum $\bm{X} \in \setC^{T \times F \times C}$. 
The time-invariant filter coefficients ${\bf g}_f \in \setC^C$ are computed as follows:
\begin{equation}
    {\bf g}_{f} = \frac{({\bf \Phi}^\text{N}_{f})^{-1}{\bf \Phi}^\text{S}_{f}}{\text{Tr}(({\bf \Phi}^\text{N}_{f})^{-1}{\bf \Phi}^\text{S}_{f})}{\bf u}
\end{equation}
where ${\bf \Phi}_f \in \setC^{C \times C}$ is the Power Spectrum Density (PSD) matrix computed both for speech (S) and noise (N) and ${\bf u} \in \setC^C$ is a one-hot vector indicating the reference channel for the beamformer, which is obtained by an attention mechanism trained jointly with the rest of the network \cite{Ochiai2017-UAF}. 
To estimate the PSD matrices, time-frequency masks ${\bf m}^\text{S\textbar N}_{t,c}$ ($\bm{\text{S\textbar N}}$ respectively for speech and noise) are estimated by two bLSTM networks jointly trained with the E2E ASR model.
The masks are estimated separately based on the magnitude spectrum of each audio channel.
Mean masks ${\bf m}^\text{S}_{t}$ and ${\bf m}^\text{N}_{t}$ are then obtained by averaging results from all the channels, and the PSD matrices are computed as follows:
\begin{equation}
    {\bf \Phi}^\text{S\textbar N}_{f} = \frac{1}{\sum^{T}_{t=1}m^\text{S\textbar N}_{t,f}}\sum^{T}_{t=1}m^\text{S\textbar N}_{t,f}{\bf x}_{t,f}{\bf x}^\dagger_{t,f} ,
\end{equation}
where $\dagger$ denotes the conjugate transpose. 
After the Neural MVDR, the enhanced power spectrogram is computed via:
\begin{equation}
    \hat{x}_{t,f} = | {\bf g}_{f}^\dagger {\bf x}_{t,f} |^2 .
\end{equation}


\subsection{SpecAugment}

\label{sec:SA}


The SpecAugment (SA) algorithm \cite{Park2019-SAA} masks (replaces by zeros) up to $F_\text{max}$ contiguous frequency bands and up to $T_\text{max}$ contiguous time frames in the ASR features $\tilde{\bf X}$.
The starting positions and the actual number of masked bands/ frames are sampled from a uniform distribution.
This process is repeated $m_F$ times for masking frequency bands and $m_T$ times for masking time frames.
To our knowledge, SA is often not used for multi-channel E2E ASR in the literature, yet we found it  very helpful for improving performance in the multi-channel scenario.

\subsection{ChannelAugment}

\label{sec:CA}


\subsubsection{Frequency Independent ChannelAugment}
In contrast to SA, the proposed ChannelAugment technique works in the complex spectrum domain rather than the ASR feature domain.
It is parameterized by a minimum ($C_\text{min}$) and a maximum ($C_\text{max}$) number of channels to keep, $0 \leq C_\text{min}, C_\text{max} < C$.
For each input example $\bm{X}$, the following algorithm is applied.
We draw a random number $C_\text{keep}$ from a uniform distribution on $\{ C_\text{min}, \dots, C_\text{max} \}$.
Then, a set $\mathcal{Z}$ of cardinality $C_\text{keep}$ is sampled from $\{1, \dots, C\}$ without replacement (\figurename~\ref{fig:CA}(a)).
Only the channels $c \in \mathcal{Z}$ are used to compute the enhanced features in the multi-channel frontend.

\noindent This general idea is applied to SF as follows. 
An augmented training example $\bm{X}' \in \setC^{T \times F \times C}$ is computed as 
\begin{equation}
x'_{t,f,c} := \left\{ \begin{array}{ll} x_{t,f,c} & c \in \mathcal{Z} \\ 0 & \text{otherwise} \end{array} \right. .
\end{equation}
According to Eq.~\eqref{eq:sf}, this is equivalent to applying the filter-and-sum operation on the reduced set $\mathcal{Z}$ of channels. 

The frequency-independent ChannelAugment can also be applied to the neural MVDR algorithm. 
We note that neural MVDR can be efficiently trained with a different set of channels in each utterance, since the weights of the mask estimator are shared across channels (unlike the SF weights).
Thus, we can compute the augmented training example by `slicing' the input tensor $\bm{X}$ so as to remove the unused channels: $\bm{X}' := \bm{X}_{:,:,\mathcal{Z}_\text{keep}}$.
In practice, we set $C_\text{min} = C_\text{max}$ in training to keep the channel dimension of $\bm{X}$ constant.
It is evident that for $C_\text{min} \ll C$, the reduced tensor dimension reduces the training time significantly.

\subsubsection{Frequency Dependent ChannelAugment}
\begin{figure}[t] 
    \centering
    \includegraphics[width=\columnwidth]{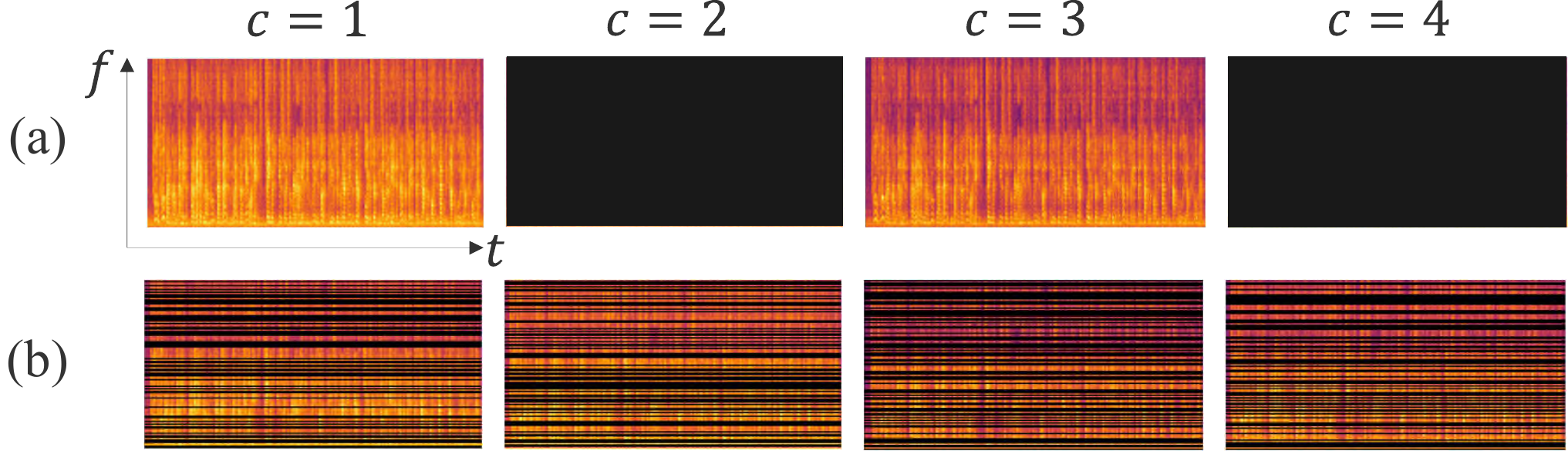}
    \caption{Visualization of ChannelAugment (as used before Spatial Filtering) on an exemplary 4-channel speech signal (showing the magnitude spectrum): (a) frequency-independent masking ($\mathcal{Z}$ = \{1,3\}); (b) frequency-dependent masking ($p_\text{keep} = 0.5$).
    }
    \label{fig:CA}
\end{figure}

For the SF algorithm, considering that it computes each frequency bin independently (cf.\ Eq.~\eqref{eq:sf}), we can also perform the channel dropping for each frequency $f$ separately.
In our work, we do so by multiplying $\bm{X}$ with a random, time-invariant dropout mask ${\bf M}$:
\begin{equation}
x'_{t,f,c} := x_{t,f,c} \cdot m_{f,c} ,
\label{eq:fdca}
\end{equation}
where $m_{f,c}$ follows a Bernoulli distribution with success probability $p_\text{keep}$.
That is, for each frequency $f$, we keep a channel $c$ with probability $p_\text{keep}$.
In this case, the number of channels kept per frequency bin follows a binomial distribution $B(C,p_\text{keep})$ rather than a uniform distribution.
\figurename~\ref{fig:CA}(b) visualizes the frequency-dependent ChannelAugment method.

\section{Data Collection}
\label{sec:datacollection}

In order to train the E2E ASR systems, we simulated multi-channel noisy and reverberant training data.
For creating the test data, we used a simultaneous playback / recording setup using two microphone arrays.

\subsection{Training Data}

The training data is based on speech from the clean training partition of Librispeech~\cite{Panayotov2015-LAA} and the English partition of Mozilla Common Voice\footnote{\url{https://commonvoice.mozilla.org/en}}. As this source data was not originally recorded in controlled environments, we applied the NISA~\cite{Sharma2020-NIE} algorithm to extract various signal characteristics and placed a threshold on estimated C50, SNR and PESQ scores to ensure that the base materials had minimal reverberation, noise and perceptual quality (25~dB for C50 and SNR and 3.0 for PESQ). The total amount of speech materials passing these criteria was 460~hrs. 
This data was artificially corrupted by convolution with 16-channel room impulse responses (RIR), generated using the Image method~\cite{Allen1979-IMF}. 
A total of 480 rooms were simulated, representing typical meeting room dimensions.
Within each room, a 16 channel Uniform Linear Array (ULA) with a 33~mm inter-mic.\ spacing was placed in 5 random positions and a directional source 
was placed in 35 random positions.
Each utterance from the source data was convolved with a randomly selected RIR, followed by addition of ambient noise. 
Microphone self noise was simulated by adding white noise,
and finally, a random gain offset was applied to each microphone to simulate inter-mic.\ gain variations. 

\subsection{Test Data}
For collecting the test data, 
a typical meeting room was setup with a USB interfaced 16 channel ULA with 33~mm inter mic spacing, mounted on a wall and an artificial mouth loudspeaker was placed in two positions (POS1: 1.8\,m, 73.5 deg.\ azimuth, POS2: 1.0\,m, 49 deg.\ azimuth). 
A subset of the Librispeech~\cite{Panayotov2015-LAA} clean test partition was used as the playback data {(}the same selection criteria was applied as used in training, to ensure only clean utterances were played back{)}. The testing partition has no overlap with training data in terms of speech or speakers. 
The recorded data was down-sampled to 16~kHz and high-pass filtered with a cut-off frequency of 50~Hz before further processing. To further demonstrate the effectiveness of our approach, the same test data was recorded with an 8 channel ULA device with the same 33~mm inter-mic.\ spacing. 

\begin{figure}[t]
    \centering
    \includegraphics[width=.9\columnwidth]{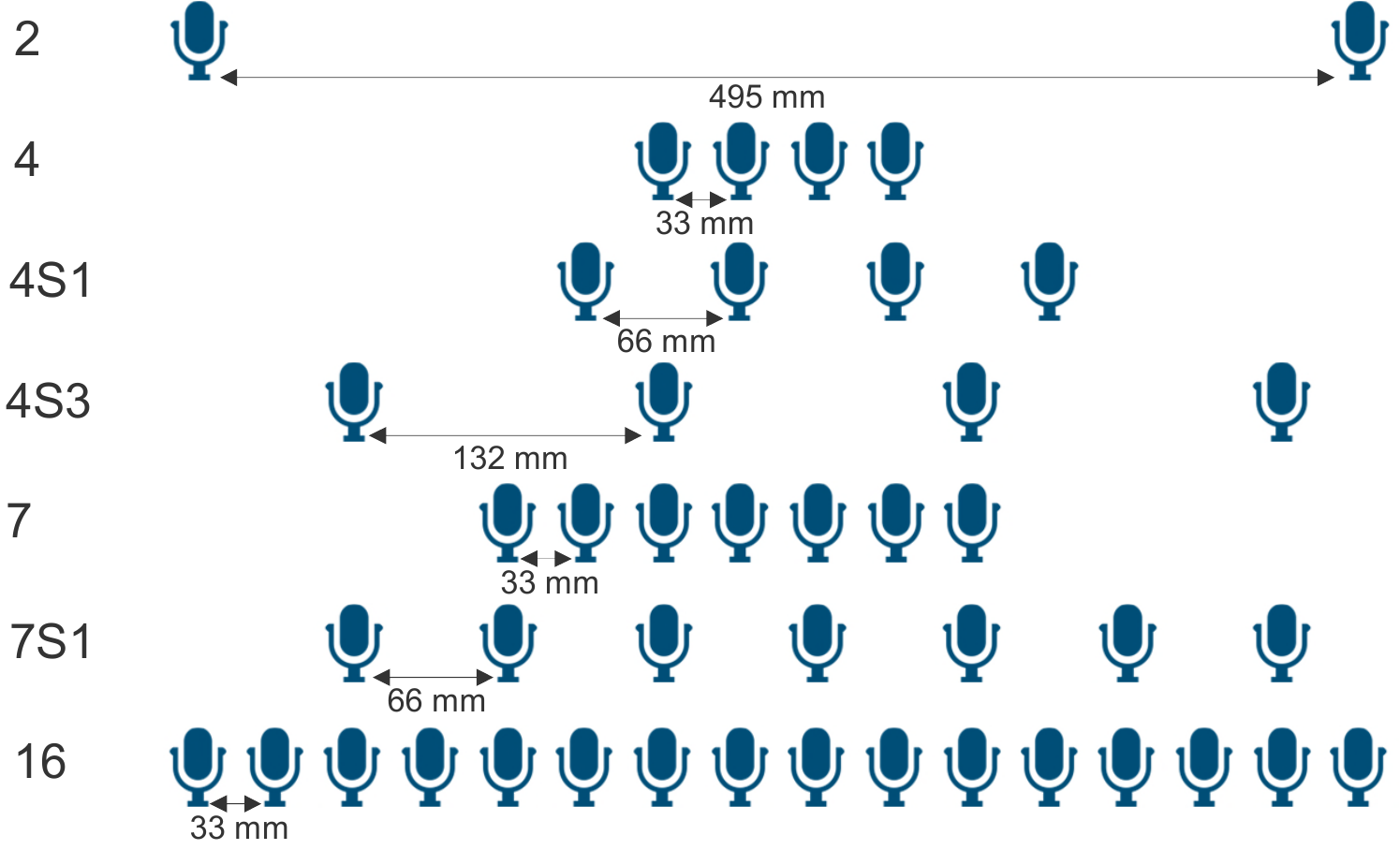}
    \caption{Linear array configurations used in testing.}
    \label{fig:mic_config}
\end{figure}

\begin{table}[t]
    \centering
    \caption{WER on the POS1 test data with models trained by applying frequency-independent ChannelAugment to SF. 
    }
    \vspace{2mm}
    \begin{adjustbox}{width=\columnwidth}
    \begin{tabular}{ll||lllllll|l}
\multicolumn{2}{l||}{WER [\%]} & \multicolumn{8}{c}{POS1 / Test configuration} \\
\hline
$C_\text{min}$ & $C_\text{max}$ & 2 & 4 & 4S1 & 4S3 & 7 & 7S1 & 16 & Avg \\
\hline\hline

\multicolumn{2}{c}{} & \multicolumn{8}{c}{No ChannelAugment} \\
\hline
-- & -- & 12.5 & 13.4 & 11.3 & 10.5 & 11.2 & 10.3 & 9.6 & 11.3
 \\
\hline
\multicolumn{2}{c}{} & \multicolumn{8}{c}{ChannelAugment} \\ 
\hline
12 & 12 & 11.7 & 12.0 & 11.0 & 10.4 & 10.7 & 10.0 & 9.6 & 10.8
 \\
4 & 4 & 11.1 & \bf 10.6 & 10.3 & 10.2 & \bf 10.1 & 10.0 & 10.1 & 10.3 
\\
12 & 16 & 11.7 & 12.2 & 11.1 & 10.5 & 10.8 & 9.8 & \bf 9.4 & 10.8 
\\
4 & 16 & 11.0 & 10.8 & \bf 10.1 & \bf 9.8 & \bf 10.1 & 9.8 & \bf 9.4 & \bf 10.1 
\\
2 & 16 & \bf 10.9 & 10.7 & 10.3 & 10.0 & 10.2 & \bf 9.7 & 9.5 & 10.2 
\\
    \end{tabular}
    \end{adjustbox}
    \label{tab:results_sfPOS1}
\end{table}

\section{Experiments and Results}
\label{sec:experiments}


\subsection{Spatial Filtering}
\label{sf}

\subsubsection{Experimental Setup}
We first performed a set of experiments training E2E multi-channel ASR models with a SF frontend.
The number of look directions for SF is set to $D=11$ and the number of input channels is $C=16$.
The encoder is composed of a VGG network with 2 stacked convolutional filters performing frame decimation with a factor of 4, then 4 bLSTM layers (size 512 for each direction) with stride 2 between the second and the third layer to obtain a total frame decimation factor of 8. The dropout rates are set to 0.4.
The decoder is composed of 2 LSTM layers with size 1024, and the dropout rates are set to 0.1 and 0.4 for the first and second layer respectively.
The training recipe is similar to \cite{Weninger2020-SSL}. 
SA is applied with $F_\text{max}=15$, $m_F=2$ in the ASR feature domain (80-dimensional log-Mel features).
We do not use time-frame masking ($T_\text{max}=m_T=0$), since we found that it led to `hallucinations' of words, which should be further investigated in the future.

For demonstrating the robustness of ChannelAugment to potential variations of the array geometry in practice, we evaluated the performance on the array configurations in \figurename~\ref{fig:mic_config}, 
which are obtained by selecting subsets of the 16-channel array microphones.

\subsubsection{Frequency Independent ChannelAugment}
\tablename~\ref{tab:results_sfPOS1} shows the results obtained on all testing configurations in \figurename~\ref{fig:mic_config} with models trained under different $C_\text{min}$ and $C_\text{max}$ values.
In the matched training/testing on 16 channels, we obtained 9.6\,\% WER on POS1.
Note that this compares to 11.7\,\% WER obtained by an E2E model trained and tested on a single channel of the array.
The configurations with increased spacing between the microphones (e.g.\ 4S1 and 4S3 vs.\ 4) generally lead to lower WER, as expected.

Without ChannelAugment, the WER for POS1 degrades by up to 40\,\% relative when training on 16 and testing on 4 channels, while ChannelAugment can reduce the degradation to 10.4\,\% relative. 
The overall best ChannelAugment configuration is to vary the number of channels between 4 and 16, where there is a slight WER improvement also in the 16 channel test case (9.6\,\% to 9.4\,\% on POS1).
In this case, the average WER is improved by 10.6\,\% relative on POS1 test data.

\subsubsection{Frequency Dependent ChannelAugment}
\begin{figure}[t]
    \centering
    \includegraphics[width=1.0\columnwidth]{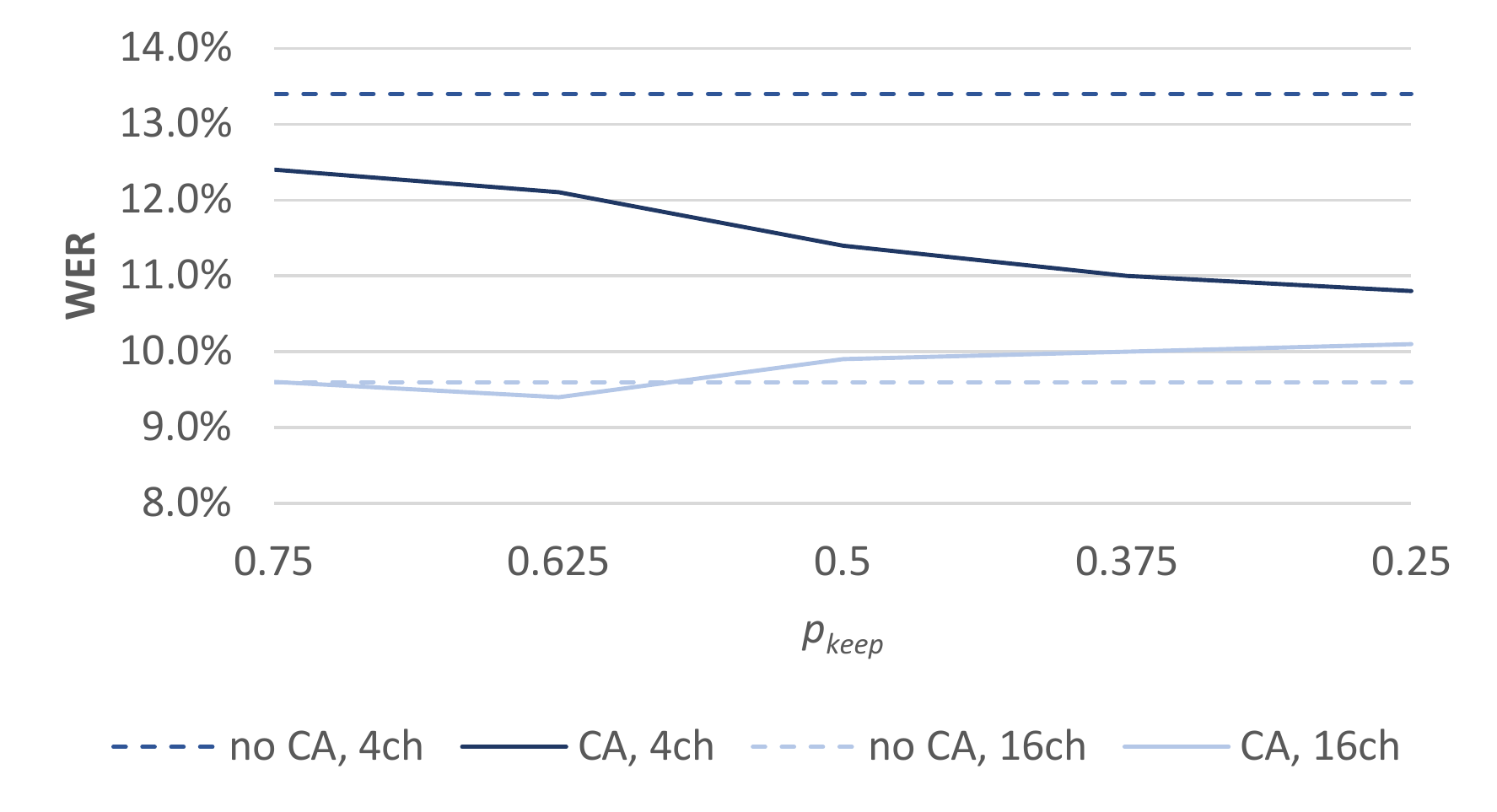}
    \caption{WER obtained by 16-channel Spatial Filtering trained with frequency-dependent ChannelAugment, using array configurations with 16 and 4 microphones (cf.\ \figurename~\ref{fig:mic_config} (b)), compared with the same configurations trained without ChannelAugment. 
    }
    \label{fig:results_sf}
\end{figure}

\figurename~\ref{fig:results_sf} shows the results with frequency-dependent ChannelAugment applied to SF, varying the keep probability of the dropout mask in training.
The evaluation is done for 16-channel and 4-channel array configurations.
As expected, the performance in the 4-channel condition improves when increasing the number of channels being dropped.
However, there is some degradation in the performance on 16 channels:
For $p_\text{keep} = 0.25$, we obtain 10.1\,\% WER in POS1, which is the same as the result with frequency-independent ChannelAugment under $C_\text{min} = C_\text{max} = 4$ (cf.\ \tablename~\ref{tab:results_sfPOS1}), but worse than the matched condition result (9.6\,\%) and the best ChannelAugment configuration from \tablename~\ref{tab:results_sfPOS1} (9.4\,\%).

\subsubsection{Additional Experiments}
\begin{table}[t]
    \centering
    \caption{WER on the POS2 test data with models trained by applying frequency-independent ChannelAugment to SF. 
    }
    \vspace{2mm}
    \begin{adjustbox}{width=\columnwidth}
    \begin{tabular}{ll||lllllll|l}
\multicolumn{2}{l||}{WER [\%]} & \multicolumn{8}{c}{POS2 / Test configuration} \\
\hline
$C_\text{min}$ & $C_\text{max}$ & 2 & 4 & 4S1 & 4S3 & 7 & 7S1 & 16 & Avg \\
\hline\hline

\multicolumn{2}{c}{} & \multicolumn{8}{c}{No ChannelAugment} \\
\hline
-- & -- & 11.8 & 12.4 & 10.7 & 9.9 & 10.8 & 9.7 & 9.2 & 10.6
 \\
\hline
\multicolumn{2}{c}{} & \multicolumn{8}{c}{ChannelAugment} \\ 
\hline
4 & 16 & 10.5 & 10.3 & 9.9 & 9.6 & 9.7 & \bf 9.3 & \bf 9.1 & 9.8
\\
2 & 16 & \bf 10.3 & \bf 10.2 & \bf 9.8 & \bf 9.5 & \bf 9.6 & 9.4 & 9.2 & \bf 9.7
\\
    \end{tabular}
    \end{adjustbox}
    \label{tab:results_sfPOS2}
\end{table}

To verify that the ChannelAugment approach is not sensitive to device location, the baseline without ChannelAugment and the two best configurations from \tablename~\ref{tab:results_sfPOS1} (frequency independent ChannelAugment, $C_\text{max} = 16$, $C_\text{min} \in \{2,4\}$) were also tested on POS2 data. 
Results are reported in \tablename~\ref{tab:results_sfPOS2}.
It can be seen that ChannelAugment still gave significant WER reduction (8.5\,\%) in this case. Note that the degradation (up to 35\,\%) under the mismatched test condition for the baseline is a bit smaller than that on the POS1 data and the WER reduction from ChannelAugment is also slightly smaller than that on the POS1 data.

The model trained with the best frequency independent ChannelAugment configuration on POS1 data ($C_\text{min} = 4$, $C_\text{max} = 16$) was also tested on the playback test data recorded through the 8-channel ULA and compared with the model trained without ChannelAugment. The WER reduction is 9.6\,\% and 4.6\,\% respectively on the recorded playback data at POS1 and POS2.

For demonstrating that ChannelAugment is complementary to the SpecAugment method and that this also holds vice versa (i.e., SpecAugment is complementary to ChannelAugment), we conducted another frequency-dependent ChannelAugment experiment at keeping probability of 0.375 with and without SpecAugment. In this case, the configuration with SpecAugment obtained an average WERR of 25\,\% (computed over all the array test configurations presented in \figurename~\ref{fig:mic_config}). Thus, we can conclude that there is little overlap between these techniques, and it is beneficial to use them simultaneously.


\subsection{Neural MVDR}

\begin{table}[t]
\caption{WER and training time per epoch (T[h]) with Neural MVDR, using fixed channel configuration (16, 4) or frequency-independent ChannelAugment with $C_\text{min} = C_\text{max} = 4$ (CA 4) for training.}
\centering
\vspace{2mm}
\begin{tabular}{l||lllll|l|l}
WER [\%] & \multicolumn{6}{c|}{POS1 / Test configuration} & T[h] \\
\hline
Train on & 2 & 4 & 4S3 & 7 & 16 & Avg & \\
\hline\hline
16 & 14.6 & 13.7 & 13.3 & 13.2 & 12.4 & 13.4 & 16.3 \\
4 & 15.4 & 14.8 & 14.1 & 14.3 & 14.1 & 14.5 & 4.7 \\
\hline
CA 4 & 14.5 & 13.9 & 13.5 & 13.3 & 13.2 & 13.7 & 4.2 \\
\end{tabular}
\vspace{-3mm}
\label{tab:results_mvdr}
\end{table}

We performed an additional proof-of-concept experiment applying ChannelAugment to the neural MVDR algorithm.
For the E2E ASR backend, we started from the CHiME-4 recipe in ESPnet \cite{Watanabe2018-EET}: the data subsampling with a factor of 4 is performed through a VGG network, then the encoder is a 4 layer bLSTM (size 1024) and finally, to mimic the SF recipe in section \ref{sf}, the decoder is composed of 2 LSTM layers (size 1024) instead of just one layer as in the original CHiME-4 recipe.
We compare the performance of training on 16 channels vs.\ training on the center 4, or on 4 randomly selected channels for the same number of epochs.
The 4 channels configuration has been chosen because it represents the best trade-off in terms of WER, training speed and GPU memory requirements. 
However, the accuracy of the fixed 4 channel training is expected to be lower than the 16 channel one, since the mask estimator of the MVDR is trained on only 1/4 of the data. 
It may well be possible that by investing more training time (i.e., running more epochs), we could regain part of this accuracy loss. 

\tablename~\ref{tab:results_mvdr} shows the results:
In the matched training and testing on 16 channels, the WER is 12.4\,\%.
Note that this number is behind the corresponding SF result mainly because the CHiME-4 recipe does not use SA.
Randomly selecting 4 channels with ChannelAugment leads to 13.2\,\% WER, while the training time is reduced by 74\,\%.
The large reduction in training time is made possible by reducing the computational complexity of the MVDR neural beamformer, which dominates the training cost since a deep BLSTM network is computed for all input channels separately.
Moreover, ChannelAugment yields a 6.4\,\% relative improvement compared to fixing the center 4 channels in training.
On average across testing configurations, the model trained with ChannelAugment performs similar to the full 16-channel training and 5.5\,\% relative better than the fixed 4-channel training.
Notably, when testing on the center 4 channels, ChannelAugment yields a sizable WER gain (14.8\,\% to 13.9\,\%)  compared to the matched training.

\section{Conclusions}
\label{sec:conclusions}

In this paper, we have introduced the ChannelAugment technique for on-the-fly data augmentation in E2E multi-channel ASR training.
Our results show that it helps to improve robustness on variants of microphone array geometry, 
and thus facilitates the deployment of E2E multi-channel ASR systems in real-world settings.
At the same time, our method works well in matched conditions, and improves training efficiency.
In future work, we will apply this technique to other front-end processing (e.g.\ \cite{Park2020-RMC}), other types of ASR models (e.g.\ \cite{Gulati2020-CCA}), and for model adaptation (cf.\ \cite{Weninger2019-LAS}). 

\bibliographystyle{IEEEbib}
\bibliography{strings,refs}

\end{document}